\documentclass[12pt,nofootinbib,superscriptaddress]{iopart}
\usepackage{graphicx}
\usepackage{iopams}

\begin{document}

\title{The Einstein static universe in Loop Quantum Cosmology}

\author{Luca Parisi$^{1,2,3}$, Marco Bruni$^{3,4,5}$, Roy Maartens$^3$\\
and Kevin Vandersloot$^3$}
 \address{$^1$Universit\`{a} di Salerno, 84081~Baronissi (Salerno),
 Italy}
\address{$^2$INFN, Sezione di Napoli, Salerno, Italy}
\address{$^3$Institute of Cosmology \& Gravitation, University of
Portsmouth, Portsmouth PO1 2EG,  UK}
\address{$^4$Dipartimento di Fisica, Universit\`{a} di Roma ``Tor Vergata", 00133 Roma,
Italy}
\address{$^5$INFN, Sezione di Roma ``Tor Vergata'', 00133 Roma, Italy}

\date{\today}

\begin{abstract}

Loop Quantum Cosmology strongly modifies the high-energy dynamics
of Friedman-Robertson-Walker models and removes the big-bang
singularity. We investigate how LQC corrections affect the
stability properties of the Einstein static universe. In General
Relativity, the Einstein static model with positive cosmological
constant $\Lambda$ is unstable to homogeneous perturbations. Using
dynamical systems methods, we show that LQC modifications can lead
to an Einstein static model which is neutrally stable for a large
enough positive value of $\Lambda$.

\end{abstract}


\maketitle

\section{Introduction}

The Einstein static universe in General Relativity (GR) is a
closed Friedman-Robertson-Walker (FRW) model that is unstable to
homogeneous perturbations~\cite{edd}. (Note that it is neutrally
stable to inhomogeneous scalar perturbations with high enough
sound speed and to vector and tensor
perturbations~\cite{Barrow:2003ni}.)

The stability of Einstein static models in high-energy
modifications of GR is an interesting mathematical question, and
is also relevant for scenarios in which the Einstein static is an
initial state for a past-eternal inflationary cosmology, the
so-called Emergent Universe scenario~\cite{Ellis:2002we}. The
standard model of cosmology (see e.g. \cite{Spergel:2006hy}) has a
flat infinite spatial geometry, and experiences a period of
primordial inflationary expansion, which is preceded by a big bang
singularity in the classical theory. Observations however do not
prove that the geometry is flat: the Universe could have nonzero
spatial curvature, as long as the late-time effect of this
curvature is very small.  In particular, a positive curvature
allows for an ``Emergent Universe" that originates asymptotically
in the past as an Einstein static universe, and then inflates and
later reheats to a hot big bang era. This model generalizes the
Eddington-Lema\^{\i}tre model~\cite{el}, and is a counter-example
to the notion that inflation can never be past-eternal and thus
cannot avoid an initial singularity -- because it is closed, it
avoids the theorems showing that inflation cannot be past
eternal~\cite{thm}.

Generalizations of the Einstein static solution in high-energy
modifications to GR have been considered in the Randall-Sundrum
braneworld scenario~\cite{Gergely:2001tn} and in $f(R)$
theories~\cite{Clifton:2005at}. The homogeneous stability of
Einstein static models in $R+\alpha R^2$ gravity has also been
investigated~\cite{bhl}.

Another theory leading to high-energy modifications of GR is Loop
Quantum Cosmology (LQC)~\cite{aa-review}, which is a canonical
quantization of homogeneous cosmological spacetimes based on Loop
Quantum Gravity~\cite{LQG-reviews}. The gravitational phase-space
variables are an su(2) valued connection and conjugate triad, and
the elementary variables underlying the quantization are the
holonomies of the connection and the fluxes of the triad. The
quantum theory obtained from LQC turns out to be inequivalent to
Wheeler-de Witt quantization (the LQC polymer representation is
different from the usual Wheeler-de Witt Schrodinger
representation). Wheeler-de Witt quantization does not resolve the
cosmological singularity, but in LQC a generic resolution of such
singularities has been obtained. Initial quantizations of LQC lead
to a regularization of the big bang
singularity~\cite{Bojowald:2001xe} resulting from the fact that
the quantum Einstein equation is non-singular as well as from
modifications to the scalar field energy density and dynamics. The
modifications to the scalar field dynamics were based on effects
arising from  quantum inverse scale factor operators.
Subsequently, the elucidation of the consequences of using
holonomies as the basic variables has shown that gravity is
modified~\cite{aps}, i.e., the structure of the Friedman equation,
rather than solely the energy density of the matter field. These
gravitational modifications typically become important at lower
energy scales than the modifications to the scalar field dynamics.

Mulryne et al.~\cite{Mulryne:2005ef} used the scalar-field
modification approach to investigate the stability of the Einstein
static model to homogeneous perturbations. They found that the new
LQC Einstein static model is a centre fixed point in phase space,
i.e. a neutrally stable point, for a massless scalar field with
$w\equiv p_\phi/\rho_\phi=1$. This modification of stability
behaviour has important consequences for the Emergent Universe
scenario, since it ameliorates the fine-tuning that arises from
the fact that the Einstein static is an unstable saddle in GR.

Here we consider the same question, but using the LQC
gravitational modifications, and neglecting the higher energy
modifications to matter. We consider a perfect fluid with
$p=w\rho$ and $w>-1$. We show that, for all $w>-1/3$, there is a
new centre, i.e. a neutrally stable point representing an Einstein
static model, but only if the cosmological constant $\Lambda$ is
above a critical scale,
 \begin{equation}
\Lambda> 6.6\pi M_P^2\,.
 \end{equation}

\section{Critical points of closed FRW models in LQC}
The loop quantum effects that we investigate manifest themselves
in the form of a modification to the classical Friedmann equation.
For the closed FRW model, the explicit form is given by \cite{aps}
\begin{equation}
H^{2}=\left(\frac{\kappa }{3}\rho +\frac{\Lambda
}{3}-\frac{1}{a^{2}} \right)\left(1-\frac{\rho }{\rho
_{c}}-\frac{\Lambda }{\kappa \rho _{c}}+ \frac{3}{\kappa \rho
_{c}a^{2}}\right),  \label{FE}
\end{equation}
where $\kappa =8\pi G=8\pi/M_P^2$, and the critical LQC energy
density is
\begin{equation}
\rho_c \approx 0.82 M_P^4\,.
\end{equation}
It is evident that the first term in parentheses is the classical
right hand side of the Friedmann equation, with the quantum
modifications appearing in the second term. The classical GR limit
is achieved in the limit as $\rho_c$ goes to infinity whence the
second term approaches unity. The classical energy conservation
equation continues to hold\footnote{Note that this means we are
not considering the inverse volume effects that would modify the
scalar field energy density and hence modify the energy
conservation equation. For the closed model there are indications
that the inverse volume effects are negligible if it is required
that the universe reach macroscopic size \cite{aps}.},
\begin{equation}
\dot{\rho }=-3H\rho \left( 1+w\right)\,.  \label{EC}
\end{equation}
 Note that
$H^2\geq0$ imposes the limits
\begin{equation}\label{lim}
{3\over a^2} \leq \kappa \rho+ \Lambda \leq \kappa \rho_c+ {3\over
a^2} \,.
\end{equation}
The modified Raychaudhuri equation follows from Eqs.~(\ref{EC})
and (\ref{FE}):
\begin{eqnarray}
\fl \dot{H}=-\frac{\kappa }{2}\rho \left( 1+w\right) \left(
1-\frac{2\rho }{
\rho_{c}}-\frac{2\Lambda }{\kappa \rho _{c}}\right)+ \nonumber \\
\!\!\! + \left[1- \frac{2\rho }{\rho_{c}}-\frac{2\Lambda }{\kappa
\rho_{c}}- \frac{3\rho (1+w)}{\rho _{c}}\right]
\frac{1}{a^{2}}+\frac{6}{\kappa \rho _{c}a^{4}}.  \label{RE}
\end{eqnarray}
We will find the critical points to the system of Eqs. (\ref{EC}),
(\ref{RE}) as well as
\begin{equation}
\dot{a}=aH\,,  \label{Hubble}
\end{equation}
which follows from the definition of $H$. The solution space is a
2-dimensional surface in the three-dimensional $(\rho, a, H)$
space, defined by the Friedman constraint~(\ref{FE}). The system
admits two critical points, which are static solutions
$\dot{a}=\dot{H}=\dot{\rho}=0$. The first critical point is the
standard GR Einstein static universe, while the second is a new
LQC Einstein static universe:
\begin{eqnarray}
&& \rho _{GR} =\frac{2\Lambda }{\kappa (1+3w)}\,,\quad
a_{GR}^{2}=\frac{2}{\kappa \rho _{GR}(1+w)}\,,  \label{STAT1} \\
&& \rho _{LQ} =\frac{2(\Lambda-\kappa \rho _{c})}{\kappa
(1+3w)}\,,\quad a_{LQ}^{2}=\frac{2}{\kappa \rho _{LQ}(1+w)}\,.
\label{STAT2}
\end{eqnarray}
The conditions under which these static solutions exist are
summarized in Table~\ref{tab1}, and follow from $a^2, \rho >0$.

A remarkable feature of the new LQ fixed point is that it is
possible to have {\em an Einstein static universe even for
vanishing cosmological constant}. Indeed, as one can see from
Eq.~(\ref{STAT2}) and Table~\ref{tab2}, when $\Lambda=0$ the LQ
fixed point exists and is unstable.

\begin{table}[htb]
\caption{\label{tab1}Conditions for the existence of Einstein
static critical points.}
\begin{indented}
\item[]\begin{tabular}{@{}*{7}{l}}
\br
 GR  & & & $\Lambda >0$ & & &$w>-1/3$  \\ \hline
      & & & $\Lambda <0$ & & &$-1<w<-1/3$ \\ \hline
LQ  & & & $\Lambda<\kappa\rho_{c}$ & & & $-1<w<-1/3$ \\ \hline
      & & & $\Lambda >\kappa\rho_{c}$ & & & $w>-1/3$ \\
\br
\end{tabular}
\end{indented}
\end{table}

For the system of equations (\ref{EC}), (\ref{RE}) and
(\ref{Hubble}), linearized stability analysis fails to give
complete information about the properties of the two critical
points, when they both exist, since in all cases, for one of the
two points there is always a pair of complex eigenvalues with
vanishing real part. In this case the linearization theorem  (see
e.g. \cite{AP92}) tells us that a fixed point which is a centre
for the linearized system is not necessarily a centre for the full
nonlinear system. In addition, the linearization of equations
(\ref{EC}), (\ref{RE}) and (\ref{Hubble}) always leads to a third
vanishing eigenvalue, simply because the actual dynamics is
two-dimensional because of the modified Friedman equation
(\ref{FE}). Therefore, using this constraint, it is convenient to
rewrite the system reducing the number of equations to two, the
dimension of the configuration space.  Solving Eq.~(\ref{FE}) for
$a^{2}$, we find that
\begin{equation}
a^2=f_\pm(\rho,H)\,,\label{2b}
\end{equation}
where
\begin{eqnarray}
f_\pm=\frac{3}{2}\frac{\left[ 2( \kappa\rho +\Lambda) +\kappa\rho
_{c}\left(\! 1\pm \sqrt{1-12H^{2}/\kappa\rho _{c}}\right)\!
\right] }{\left( \kappa\rho +\Lambda \right) ^{2}+ \kappa \rho
_{c}\left( 3H^{2}-\kappa\rho -\Lambda\right)}.   \label{aFE1}
\end{eqnarray}
Substituting this into Eq.~(\ref{RE}), we find two branches for
the time derivative of the Hubble parameter,
\begin{eqnarray}
\mbox{LQ} &:&\dot{\rho}=-3H\rho \left( 1+w\right) \quad \mbox{and}
\quad
\dot{H}=F_{+}(\rho ,H)\,,\label{lq}\\
\mbox{GR} &:&\dot{\rho}=-3H\rho \left( 1+w\right) \quad \mbox{and}
\quad \dot{H}=F_{-}(\rho ,H)\,,\label{gr}
\end{eqnarray}
where
\begin{eqnarray}
\fl F_{\pm}=-\frac{\kappa }{2}( 1+w)\rho \left( 1-\frac{2\rho }{
\rho_{c}}-\frac{2\Lambda }{\kappa \rho _{c}}\right)+
{6\over \kappa\rho_c f_\pm^2} +  \nonumber \\
+ {1\over f_\pm}\left[1-{2\rho\over \rho_c }-{2\Lambda
\over\kappa\rho_c }- {3(1+w)}{\rho\over \rho_c}\right].
  \label{f1}
\end{eqnarray}

In the classical limit,
\begin{eqnarray}
\lim_{\rho _{c}\rightarrow \infty }f_{+}&=&0\,,\label{a1} \\
\lim_{\rho _{c}\rightarrow \infty }f_{-} &=&\frac{3}{\kappa\rho
-3H^{2}+\Lambda }\,,
\end{eqnarray}
where the second equation is the GR Friedman equation. This shows
how the two branches in Eq.~(\ref{2b}) recover the GR and the new
quantum static solution. In addition,
\begin{eqnarray}
\lim_{\rho _{c}\rightarrow \infty }F_{+} &=&\infty\,, \\
\lim_{\rho _{c}\rightarrow \infty }F_{-}
&=&-H^{2}-\frac{\kappa\rho }{6}(1+w)+ \frac{\Lambda }{3}\,,
\end{eqnarray}
where the first limit is consistent with Eq.~(\ref{a1}), and the
second limit gives the GR Raychaudhuri equation.

The system (\ref{lq}) admits the static solution with $\rho_{LQ}$
as in Eq.~(\ref{STAT2}). Substituting this into Eq.~(\ref{aFE1})
reproduces $a^2_{LQ}$ as in Eq.~(\ref{STAT2}). We evaluate the
eigenvalues of the Jacobian matrix at this point, to find the two
eigenvalues
\begin{equation}
\lambda _{LQ}=\pm\sqrt{(\kappa\rho_{c}-\Lambda)(1+w)}\,.
\end{equation}
Thus the LQ fixed point is either unstable (of the saddle kind),
when $\kappa\rho_{c}>\Lambda$ and $-1<w<-1/3$, or a centre for the
linearized system, i.e. a neutrally stable fixed point, when
$\kappa\rho_{c}<\Lambda$ and $w>-1/3$. (The limits take into
account the conditions in Table~\ref{tab1}.) Again, for the latter
point the linearized analysis is not sufficient and therefore we
turn to a numerical analysis in the next section.

For the system Eq.~(\ref{gr}), we find the GR static solution, and
the eigenvalues of the linearized system are
\begin{equation}
\lambda _{GR}=\pm\sqrt{\Lambda(1+w)}\,.
\end{equation}
These are real with opposite signs for $\Lambda>0$ and $w>-1/3$,
so that the fixed point is unstable (of the saddle type). For
$\Lambda<0$ and $-1<w<-1/3$, the fixed point is a centre, as
confirmed by numerical analysis. It does not appear to be widely
known that the GR Einstein static universe can be neutrally stable
when $\Lambda<0$ (see also~\cite{bhl}).

The results of the linearized stability analysis are summarized in
Table~\ref{tab2}.

\begin{table}[htb]
\caption{\label{tab2}Eigenvalues for the critical points in
Table~\ref{tab1}.}
\begin{indented}
\item[]\begin{tabular}{@{}*{5}{l}}
\br & $\lambda_{1}$ & $\lambda_{2}$ \\ \hline
GR &  &  \\
$\Lambda >0$ \quad and \quad $w>-1/3$    &    $>0$ & $<0$ \\
$\Lambda <0$ \quad and \quad $-1<w<-1/3$
& $\mbox{Re}(\lambda_{1})=0$ & $\mbox{Re}(\lambda_{2})=0$ \\
& $\mbox{Im}(\lambda_{1})>0$ & $\mbox{Im}(\lambda_{2})<0$\\ \hline
LQ &  &  \\
$\Lambda<\kappa\rho_{c}$ \quad and\quad $-1<w<-1/3$ &
$>0$ & $<0$ \\
$\Lambda>\kappa\rho_{c}$ \quad and \quad $w>-1/3$
& $\mbox{Re}(\lambda_{1})=0$ & $\mbox{Re}(\lambda_{2})=0$ \\
& $\mbox{Im}(\lambda_{1})>0$ & $\mbox{Im}(\lambda_{2})<0$ \\
 \br
\end{tabular}
\end{indented}
\end{table}

\section{Numerical integration}

In order to extend the linearized stability analysis, we perform
numerical integrations of the systems~(\ref{lq}) and (\ref{gr}).
We also integrate the nonlinear system of Eqs.~(\ref{EC}),
(\ref{RE}) and (\ref{Hubble}), with initial conditions fulfilling
the Friedman constraint (\ref{FE}), in order to show the full
configuration space diagrams and gain a better understanding of
the dynamics. We take $\Lambda>0$ and $w=1$.

\begin{figure}[h]
\begin{center}
\includegraphics*[scale=.40,angle=-90]{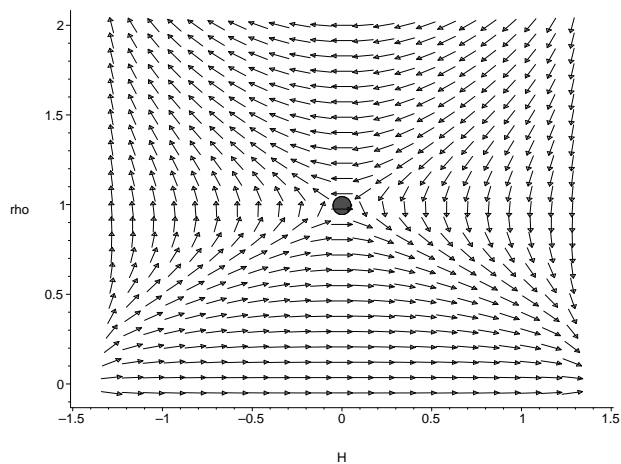}
\end{center}
\caption{Dynamical behaviour of the system around the GR fixed
point for the case $\Lambda>\kappa\rho_{c}$, with
$\Lambda/\kappa=2$, $w=1$ (using units $M_P=1$).} \label{1}
\end{figure}

\begin{figure}[h]
\begin{center}
\includegraphics*[scale=.38]{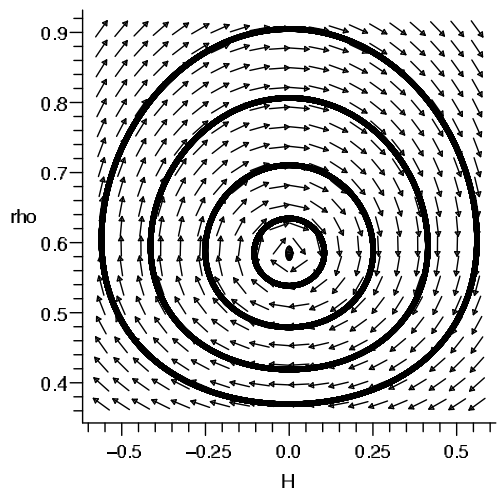}
\end{center}
\caption{As in Fig.~\ref{1}, for the LQ fixed point.} \label{2}
\end{figure}

First we consider the case $\Lambda>\kappa\rho_{c}$. The
$(H,\rho)$ plots in Figs.~\ref{1} and \ref{2} are obtained by
integrating the system~(\ref{lq}) and (\ref{gr}) in some
neighbourhood of the fixed points. The plots show that the GR
Einstein static solution is a saddle while the LQ Einstein static
solution is a centre.

\begin{figure}[h]
\begin{center}
\includegraphics*[scale=.45]{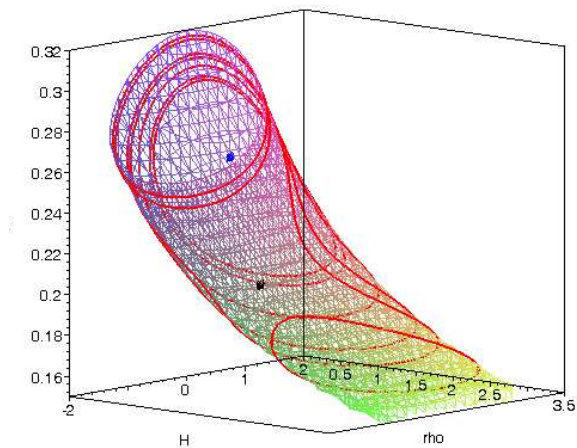}
\end{center}
\caption{Trajectories on the Friedman constraint surface, for the
same parameters as in Figs.~\ref{1} and \ref{2}. The GR fixed
point is at the bottom of the Friedman surface, while the LQ point
is at the top. Note that some trajectories wrap around the ``tube"
but cannot be shrunk continuously to the LQ fixed point.}
\label{3}
\end{figure}

A better understanding can be obtained by plotting the whole 3D
space $(H,\rho,a)$ for a wide range of initial conditions, shown
in Fig.~\ref{3}. The trajectories lie on the 2-dimensional
Friedman constraint surface.

\begin{figure}[h]
\begin{center}
\includegraphics*[scale=.40]{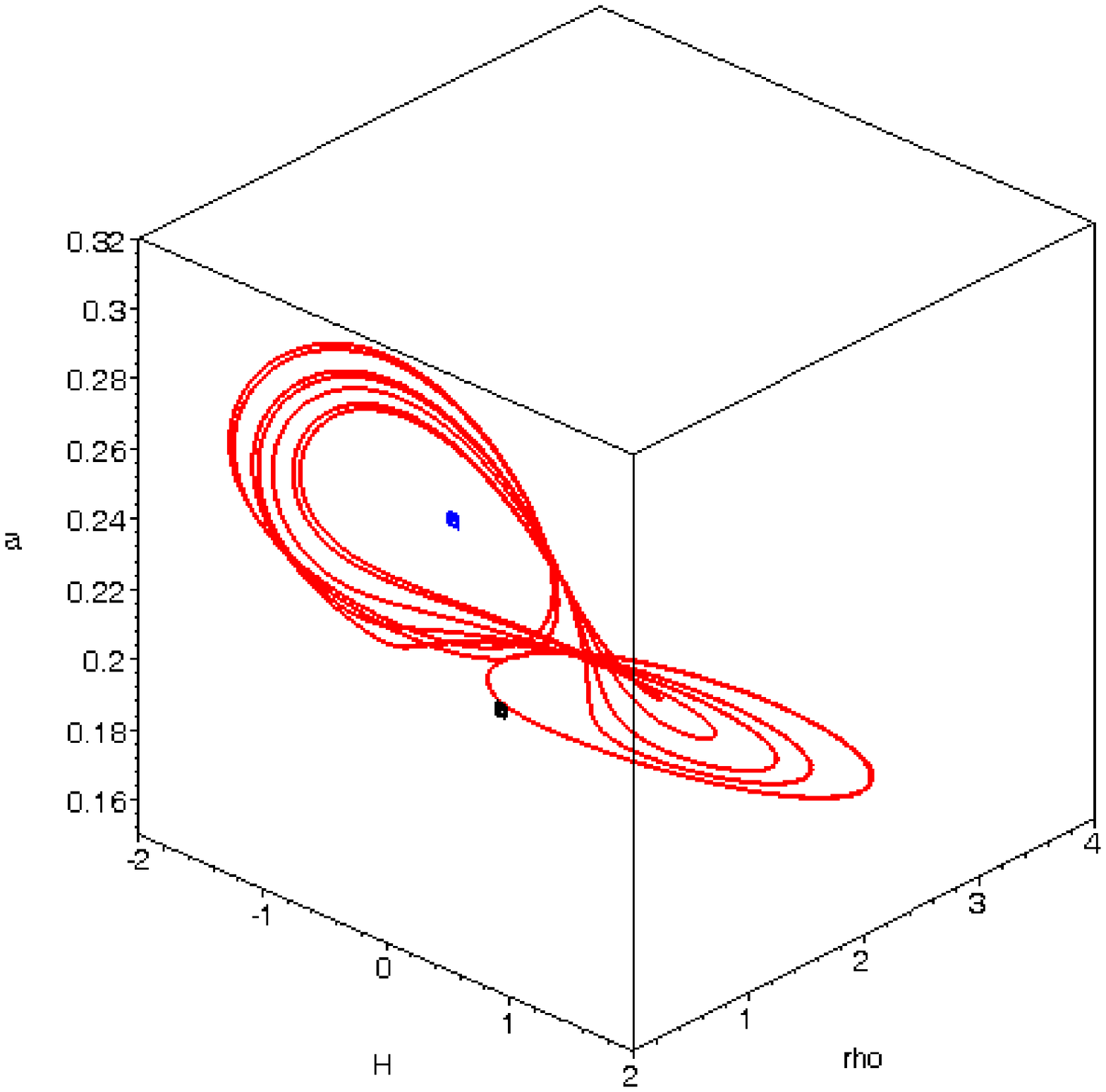}
\end{center}
\caption{As in Fig.~\ref{3}, showing trajectories that start far
from the fixed points.} \label{4}
\end{figure}

The behaviour near the fixed point is in agreement with the
linearized stability analysis, but new interesting features arise.
For initial conditions far enough from the fixed point, there are
trajectories that wrap around the Friedman tube, so that cyclic
models are possible even if they are not related with the centre
fixed point, since they cannot be shrunk to a point; see
Fig.~\ref{4}.

\begin{figure}[h]
\begin{center}
\includegraphics*[scale=.32]{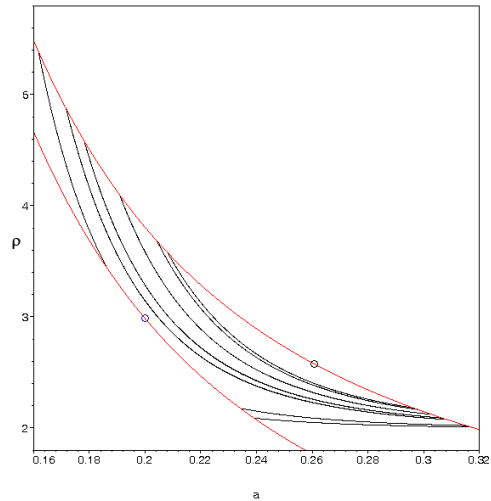}
\end{center}
\caption{The upper red (grey) curve is $\rho_{+}(a)$ and
$\rho_{-}(a)$ is the lower curve. Trajectories for several initial
conditions are depicted in black. The GR fixed point is the box on
the curve $\rho_{-}$, and the LQ point is the circle on the curve
$\rho_{+}$. } \label{5}
\end{figure}

This behaviour can also be interpreted via a plot of the energy
density against $a$. Defining
\begin{equation}
\rho _{-} =\frac{3}{\kappa a^{2}}\,,~
\rho_{+}=\rho_{c}+\frac{3}{\kappa a^{2}}\,,~\rho_{m}=\rho +
{\Lambda \over \kappa}\,,   \label{rho12}
\end{equation}
Eq.~(\ref{FE}) can be written in the form
\begin{equation}
H^{2}= \frac{\kappa }{3\rho_{c}} \left(\rho_{m} - \rho_- \right)
\left( \rho_{+} - \rho_{m} \right). \label{FEnew}
\end{equation}
When $\rho_{m}=\rho _{-}(a)$ or $\rho_{+}(a)$, then $H$=0, so the
system undergoes a bounce or starts a recollapsing phase,
respectively. This is illustrated in Fig.~\ref{5}.

\begin{figure}[h]
\begin{center}
\includegraphics*[scale=.30]{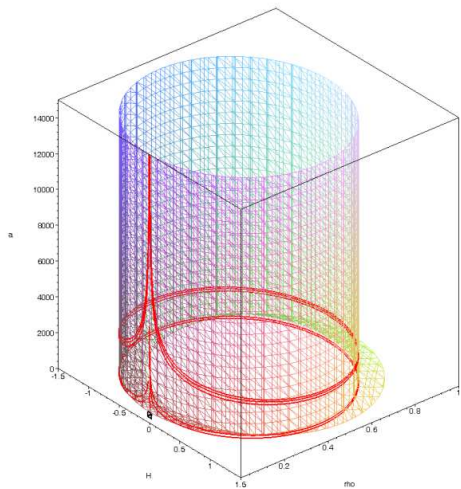}
\end{center}
\caption{The case $\Lambda/\kappa=4\times10^{-7}$, with $w=1$. }
\label{6}
\end{figure}

The second case $\Lambda<\kappa\rho_{c}$ is illustrated in
Fig.~\ref{6}. For $w=1$, only the GR fixed point is present and it
is unstable. The trajectories are again wrapped around the
Friedman surface. During a contracting phase they tend to a
minimum of the scale factor and then, after a bounce, there is a
phase of expansion.

We can integrate Eq.~(\ref{EC}) for $\rho$ as a function of $a$,
and substitute the expression for $a(\rho)$ into the modified
Friedman equation~(\ref{FE}). Using the dimensionless variables,
\begin{eqnarray*}
x^{2}= \frac{3\kappa\rho_{c}}{\Lambda^{2}}H^{2}\,,~~
y=\frac{\kappa\rho}{\Lambda}\,,~~ B=
\frac{\kappa\rho_{c}}{\Lambda}\,, \label{newvars}
\end{eqnarray*}
this leads to
\begin{equation}
x^{2}= \!\left[\! y+1-Cy^{2/3(w+1)} \right]\!\left[\!B-
y-1+Cy^{2/3(w+1)} \right]\!, \label{NFE}
\end{equation}
where $C$ is a constant of integration. This expression allows us
to get a better understanding of the particular case depicted in
Figs.~\ref{3} and \ref{4}, where both the fixed points are
present, by finding the separatrix. The separatrix is a curve
(actually, a union of orbits) which marks the boundary of regions
where the dynamical behaviour of the system is different. In this
case the separatrix is the junction of the stable and unstable
manifold with the GR Einstein static hyperbolic fixed point. First
we solve Eq.~(\ref{NFE}) for $C$, then we substitute numerical
values for the parameters $w=1$ and $B=.4121769562$ (using
$M_P=1$), and we also substitute $x=0$ and $y=1/2$ for the GR
fixed point. This produces a numerical value for $C$. Through this
procedure, the equation of the separatrix is implicitly given by
Eq.~(\ref{NFE}) with fixed values of the parameters ${C}$, $w$ and
$B$.

\begin{figure}[h]
\begin{center}
\includegraphics*[scale=.45]{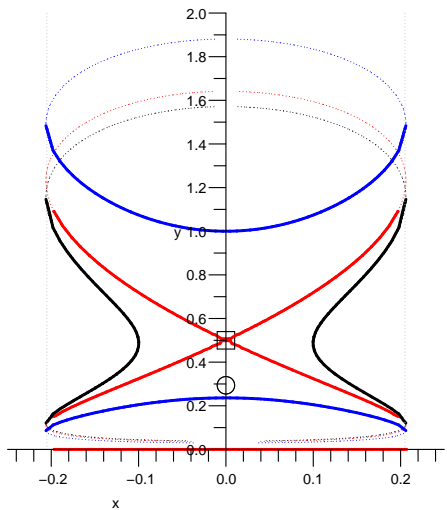}
\end{center}
\caption{Projection of the phase space onto the $(x,y)$-plane for
the case of Figs.~\ref{3} and \ref{4}. It represents the
trajectories as seen from the bottom of the Friedman surface. Both
the GR fixed point (a box) and the LQ fixed point (a circle)
exist. The two vertical lines are the edges of the Friedman
surface. The red curve represents the two branches of the
separatrix.} \label{7}
\end{figure}

More insight can be acquired via the plot of the separatrix
projected onto the $(x,y)$-plane, as shown in Fig.~\ref{7}. The
curves depicted are the separatrix, which wraps around the
Friedman surface, and some other trajectories which are first
integrals obtained for different values of the integration
constants. The closed loops around the Friedman surface are
outside the region marked by the separatrix; the other curve is
not closed around the tube, and it can be shrunk continuously to
the LQ fixed point.

\section{Conclusions}

We have shown that LQC gravitational modifications to the Friedman
equation lead to a new high-energy critical point for the Einstein
static universe. This LQC Einstein static model is an unstable
saddle (like the standard GR Einstein static solution) for
sufficiently negative pressure and a sub-critical cosmological
constant, i.e.,
\begin{equation}
-1<w<-{1\over3}\,,~~~ \Lambda< \kappa\rho_c\,.
\end{equation}
If $w$ is large enough and the cosmological constant is above the
critical value, then the LQC Einstein static model is a centre:
\begin{equation}\label{cs}
w>-{1\over3}\,,~~~ \Lambda> \kappa\rho_c\,.
\end{equation}
This neutrally stable behaviour is in strong contrast to the GR
case, where the Einstein static model is unstable for all
(positive) values of $\Lambda$.

Modified stability of the Einstein static model is also found for
$R+\alpha R^2$ gravity~\cite{bhl} and for the LQC case with matter
modifications, rather than gravitational
modifications~\cite{Mulryne:2005ef}. This illustrates the general
point that high-energy modifications to GR, which typically
strongly modify the big bang singularity, also modify the
dynamical nature of the non-singular Einstein static universe.

Our result means that the fine-tuning problem for the Emergent
Universe scenario is qualitatively changed by LQC gravitational
modifications. In GR, the Einstein static is an unstable saddle,
so that severe fine-tuning is required if the Einstein static is
to be the initial state for a past-eternal inflationary cosmology.
With LQC gravitational modifications, the initial state becomes a
centre when Eq.~(\ref{cs}) holds, thus ameliorating the
fine-tuning. The same point was made in Mulryne et
al.~\cite{Mulryne:2005ef} in the case of LQC matter modifications.
In that case however, the analysis is restricted to a massless
scalar field ($w=1$), and the centre exists for all $\Lambda$. Our
analysis applies for all perfect fluids with $w>-1$, since we
focus on the LQC gravitational modifications. These gravitational
modifications operate typically at a lower energy scale than the
matter modifications. Clearly, from the standpoint of the Emergent
Universe scenario, one needs a further mechanism to break the
infinite cycles about the centre fixed point, leading to a
subsequent inflationary phase as discussed in \cite{Lidsey:2006md}
for a general class of braneworld models sourced by a scalar field
with a constant potential. This mechanism will be the subject of
future work.

\ack LP would like to thank the ICG, Portsmouth for warm
hospitality and friendship, and in particular Christian B\"ohmer,
Giuseppe De Risi, Francisco Lobo and Sanjeev Seahra for
enlightening discussions. LP was partially supported by PRIN 2006
of the Italian Ministero dell'Universit\`a e della Ricerca. MB
thanks the MIUR "Rientro dei Cervelli" scheme for funding while
this research was carried out. The work of RM is partly supported
by PPARC/STFC. KV is supported by Marie Curie Incoming
International Fellowship MIF1-CT-2006-022239.

\end{document}